\documentclass[aps,prx,showpacs,twocolumn,superscriptaddress,longbibliography]{revtex4-1}

\usepackage[english]{babel}

\usepackage{graphicx}
\usepackage{amsmath}
\usepackage{multirow}
\usepackage{tikz}
\usetikzlibrary{calc}
\usepackage{hyperref}

\newcommand\tikzmark[1]{\tikz[remember picture,overlay] \coordinate (#1);}

\def\aap{Astron. Astrophys.} 
\def\araa{ARA\&A}             
\def\procspie{Proc.~SPIE}   
\def\mnras{Mon. Not. R. Astron. Soc.}           
\def\pasp{Publ. Astron. Soc. Pac.}               

\begin{document}


\title{Detection and Implications of Laser-Induced Raman Scattering at Astronomical Observatories}


\author{Fr\'ed\'eric P.~A. Vogt}
\email[]{fvogt@eso.org}
\thanks{ESO Fellow}
\affiliation{European Southern Observatory (ESO), Avenida Alonso de C\'ordova 3107, 763 0355 Vitacura, Santiago, Chile.}

\author{Domenico Bonaccini Calia} 
\author{Wolfgang Hackenberg}       

\affiliation{European Southern Observatory (ESO), Karl-Schwarzschild-Stra{\ss}e 2, 85748 Garching, Germany.}

\author{Cyrielle Opitom}                 
\thanks{ESO Fellow}
\affiliation{European Southern Observatory (ESO), Avenida Alonso de C\'ordova 3107, 763 0355 Vitacura, Santiago, Chile.}

\author{Mauro Comin}                    

\affiliation{European Southern Observatory (ESO), Karl-Schwarzschild-Stra{\ss}e 2, 85748 Garching, Germany.}

\author{Linda Schmidtobreik}            
\author{Jonathan Smoker}                 
\author{Israel Blanchard}                    
\author{Marcela Espinoza Contreras} 
\author{Ivan Aranda}                            
\author{Julien Milli}                              
\author{Yara L. Jaffe}                           
\thanks{ESO Fellow}
\author{Fernando Selman}                  
\author{Johann Kolb}                            
\author{Pascale Hibon}                        

\affiliation{European Southern Observatory (ESO), Avenida Alonso de C\'ordova 3107, 763 0355 Vitacura, Santiago, Chile.}

\author{Harald Kuntschner}          
\author{Pierre-Yves Madec}          

\affiliation{European Southern Observatory (ESO), Karl-Schwarzschild-Stra{\ss}e 2, 85748 Garching, Germany.}


\date{\today}

\begin{abstract}
Laser guide stars employed at astronomical observatories provide artificial wavefront reference sources to help correct (in part) the impact of atmospheric turbulence on astrophysical observations. Following the recent commissioning of the 4 Laser Guide Star Facility (4LGSF) on the Unit Telescope 4 (UT4) of the Very Large Telescope (VLT), we characterize the spectral signature of the uplink beams from the 22~W lasers to assess the impact of laser scattering from the 4LGSF on science observations. We use the Multi-Unit Spectroscopic Explorer (MUSE) optical integral field spectrograph mounted on the Nasmyth B focus of UT4 to acquire spectra at a resolution of R$\cong$3000 of the uplink laser beams over the wavelength range of 4750\,\AA\ $\rightarrow$ 9350\,\AA. We report the first detection of laser-induced Raman scattering by N$_2$, O$_2$, CO$_2$, H$_2$O and (tentatively) CH$_4$ molecules in the atmosphere above the astronomical observatory of Cerro Paranal. In particular, our observations reveal the characteristic spectral signature of laser photons -- but 480\,\AA\ to 2210\,\AA\ redder than the original laser wavelength of 5889.959\,\AA\ -- landing on the 8.2m primary mirror of UT4 after being Raman-scattered on their way up to the sodium layer. Laser-induced Raman scattering, a phenomenon not usually discussed in the astronomical context, is not unique to the observatory of Cerro Paranal, but common to any astronomical telescope employing a laser-guide-star (LGS) system. It is thus essential for any optical spectrograph coupled to a LGS system to handle thoroughly the possibility of a \textit{Raman spectral contamination} via a proper baffling of the instrument and suitable calibrations procedures. These considerations are particularly applicable for the HARMONI optical spectrograph on the upcoming Extremely Large Telescope (ELT). At sites hosting multiple telescopes, laser collision prediction tools also ought to account for the presence of Raman emission from the uplink laser beam(s) to avoid the unintentional contamination of observations acquired with telescopes in the vicinity of a LGS system.\\

\textbf{Popular summary:} Atmospheric turbulence strongly affects the sharpness of astronomical observations from the ground. Specially-equipped telescopes (and their associated instruments) can reduce this effect by directing lasers up into the sky, causing sodium atoms in the upper atmosphere to glow. Deformable mirrors can then use these "artificial guide stars" to help correct for the impact of turbulence on observations. Four such lasers were recently installed at the Very Large Telescope (VLT) at Cerro Paranal in Chile. For the first time, we characterized the astronomical consequences of laser-induced inelastic Raman scattering, a process through which the laser photons lose energy by exciting air molecules. This is a possible source of contamination for astrophysical observations, appearing in the data as complex groups of emission lines.\\

We used the Multi-Unit Spectroscopic Explorer (MUSE) integral field spectrograph -- an instrument that obtains a spectrum for each pixel of an image in a given area of the sky -- on the Unit 4 Telescope of the VLT to record the spectral signature of the lasers over the entire optical range. These 22-watt lasers, each tuned to a wavelength of 5889.959\,\AA, excite sodium atoms at 90\,km above the ground. We identified contaminating Raman spectral lines from molecular nitrogen, molecular oxygen, carbon dioxide, water, and tentatively methane. This detailed characterization of the spectral signature of laser-induced Raman scattering, and the identification of the molecules involved, is crucial to anticipating, reducing and correcting the possible contamination of scientific observations obtained at any observatory using a laser guide star.
\end{abstract}


\maketitle

$ $\\
$ $\\
\section{Introduction}

The past few years have seen the emergence of numerous Laser Guide Star (LGS) systems at astronomical observatories \citep[][]{Liu2006,Liu2008}. These provide artificial wavefront reference sources to help alleviate the impact of atmospheric turbulence on astronomical observations \citep[][]{Foy1985,Davies2012}. A cleaner wavefront does not only lead to an enhancement in image quality: it also improves the detection limit of the telescope by further concentrating the light of point sources, a crucial advantage for the detection (from the ground) of galaxies at the highest redshifts, for example. Most of the existing LGS systems are available to the general community of observers, and thus used regularly in operations \citep[][]{dOrgeville2016}. Such is the case at the W.M. Keck telescopes \citep{Wizinowich2006, vanDam2006}, the Gemini North and South telescopes \citep{Boccas2006,Christou2010,dOrgeville2012,Rigaut2014,Neichel2014}, the Subaru telescope \citep{Hayano2010,Minowa2012} and the Very Large Telescope (VLT) \citep{Lewis2014}.

The 4 Laser Guide Star Facility \citep[4LGSF;][]{BonacciniCalia2014} consists of four laser guide star units (LGSUs), each comprised of a standalone 22 W laser \citep{Feng2009,Taylor2009,BonacciniCalia2010}, a launch telescope, steering mechanisms and dedicated control electronics. These systems are an integral part of the Adaptive Optics Facility \citep[AOF;][]{Arsenault2013} currently being installed on the Unit Telescope 4 (UT4) of the VLT on Cerro Paranal in Chile. In addition to the 4LGSF, the AOF also comprises a new deformable secondary mirror \citep{Arsenault2006,Briguglio2014} and the adaptive optics modules GALACSI \citep{Stuik2006,LaPenna2016} \& GRAAL \citep{Paufique2010,Paufique2016,Hibon2016}. 

The AOF is intended to feed corrected wavefronts to the MUSE optical integral field spectrograph \citep[spectral range: 4750\,\AA\ $\rightarrow$ 9350\,\AA;][]{Bacon2010} and the HAWK-I near-IR wide-field imager \citep{Kissler-Patig2008,Siebenmorgen2011}. Altogether, these systems allow to correct \textit{on-the-fly} the effect of atmospheric turbulence over several arcminutes using a square asterism of 4 LGSs, the spatial extent of which is adjustable to the respective field-of-views of MUSE (7.5 arcseconds in Narrow-Field-Mode$\equiv$NFM; 60 arcseconds in Wide-Field-Mode$\equiv$WFM) and HAWK-I (7.5 arcminutes). Specifically, the adaptive optics corrections to be provided by the AOF are a) laser tomography for the MUSE NFM, and b) ground-layer correction for the MUSE WFM and HAWK-I \citep{Arsenault2006a}. 

As the power of LGS systems increase, so is their potential for collateral damage on the associated astrophysical observations. In the case of Cerro Paranal, the 22~W of the 4LGSF lasers represent a factor of four increase in the on-sky power compared to the PARLA laser \citep[][]{Lewis2014} used with the SINFONI infrared integral field spectrograph \citep{Eisenhauer2003,Bonnet2004} on UT4 since March 2013. This power increase, the fact that the 4LGSF lasers will be used (at times) with an optical spectrograph, and the presence of numerous telescopes in the immediate vicinity of UT4 on Cerro Paranal motivated the spectral analysis of the 4LGSF uplink beams presented in this Article. In particular, the present analysis distinguishes itself from previous observations targeting LGS systems \citep[][]{Hayano2003,Coulson2010,Amico2015} in that 1) it does not restrict itself to the immediate spectral vicinity of the laser wavelength, and 2) it focuses on the existence and consequences of inelastic Raman scattering physics (in addition to elastic Rayleigh and Mie scattering) associated with the use of LGS systems, a phenomena hitherto often overlooked within the astrophysics community.

This  Article is structured as follows. Our initial detection of laser-induced Raman scattered photons by N$_2$ and O$_2$ molecules above Cerro Paranal is described in Sect.~\ref{sec:first}. We discuss our search for the signatures of Raman scattering from additional molecules in Sec.~\ref{sec:more}, and quantify the fluxes of the different Raman lines in Sec.~\ref{sec:fluxes}. We conclude by discussing the impact of laser-induced Raman scattering physics on astrophysical observations in Sec.~\ref{sec:astro}. Unless mentioned otherwise, all wavelengths are quoted in air at a pressure of 1 atm and a temperature of 15$^{\circ}$C.

\section{The first detection of laser-induced atmospheric Raman scattering at Cerro Paranal}\label{sec:first}

\begin{figure*}
\centerline{\includegraphics[width=\textwidth]{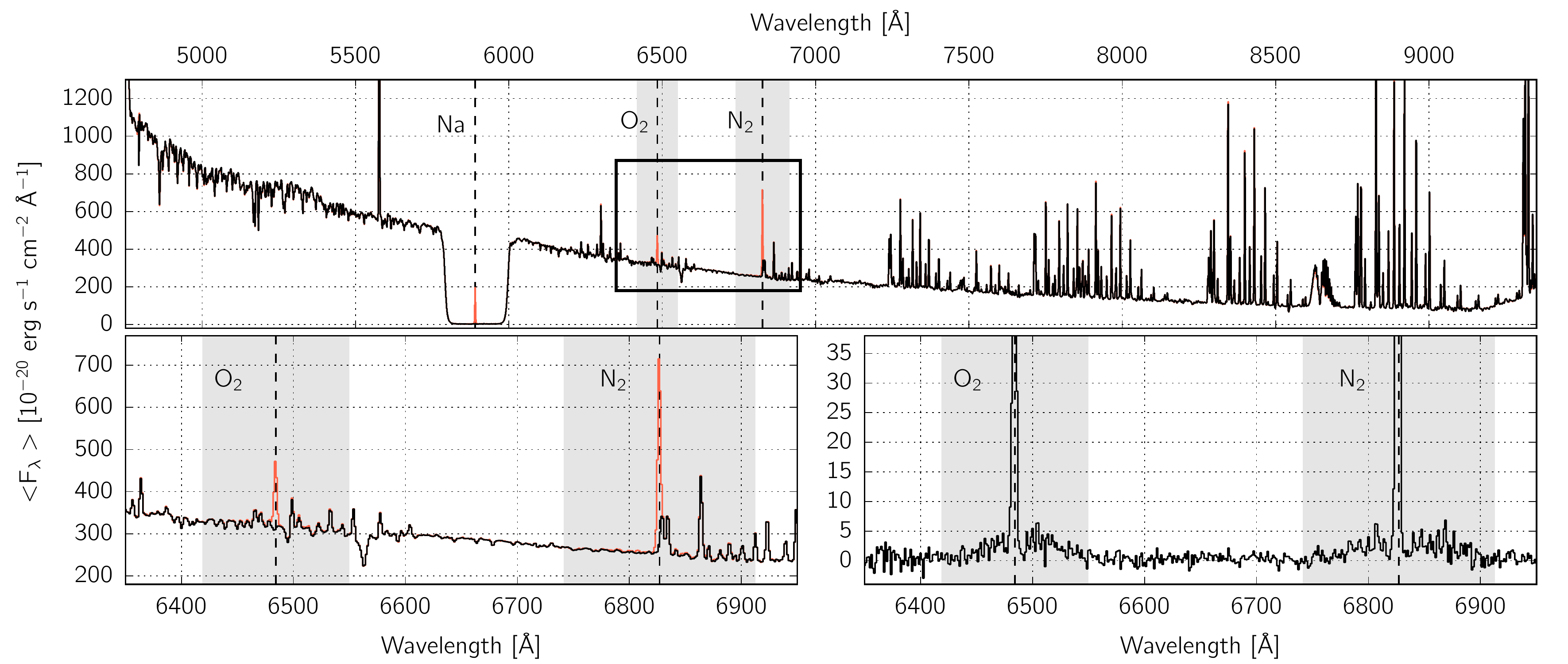}}
\caption{Top: mean MUSE spectra (in the WFM-Nominal$\equiv$WFM-N setup) constructed from a 60\,s exposure on an empty sky field at airmass $\sim$1.8, with and without the 4LGSF lasers propagating (red and black curves, respectively), under full Moon conditions. In addition to the residual laser line (visible through Rayleigh and Mie scattering in the laser beams) at $\lambda_\mathrm{4LGSF}$ (labelled ``Na''), spectral features resulting from the inelastic Raman scattering of laser photons by the O$_2$ and N$_2$ photons are present at 6484.39\,\AA\ and 6827.17\,\AA\ (respectively) when the 4LGSF lasers are propagating. Bottom left: idem, but zooming-in on the 6350\,\AA\ $\rightarrow$ 6950\,\AA\ spectral range. Bottom right: difference between the spectra of the top panel, revealing the rotational wings (i.e. the $S$-branch and $O$-branch) surrounding the central vibrational peak (i.e. $Q$-branch) of the Raman lines for both the O$_2$ and N$_2$ molecules. The spectral extent of the Raman contamination (including both the vibrational central peak and rotational side-bands) is traced using greyed-out areas in all panels. The box in the top panel shows the extent of the bottom panels.}\label{fig:nominal}
\end{figure*}

We undertook dedicated observations combining the newly installed 4LGSF systems with the MUSE integral field spectrograph (located on the Nasmyth B focus of UT4) in 2016, June, November \& December, as part of the 4LGSF/AOF ongoing commissioning activities. These observations were designed to derive an initial characterization of the possible impact of laser scattering from the 4LGSF on science observations. We note that these observations were performed \emph{prior} to the installation of the GALACSI AO module (that has since then been inserted directly ahead of MUSE in the light path). All observations were reduced using biases, flat fields (including both lamp and sky flats) and arc lamp exposures acquired as part of the regular calibration plan of MUSE. The data were flux-calibrated using dedicated observations of flux standard stars (one per night) taken from the MUSE list of standard stars.

At the laser wavelength, the laser uplink beams are visible through elastic, non-isotropic, polarization-preserving Rayleigh scattering of the laser photons by the air molecules, and through non-isotropic, higher intensity Mie scattering by dust particles. In the specific case of the 4LGSF, 18\,W are emitted by each LGSU at the laser main carrier wavelength $\lambda_\text{4LGSF,vacuum}=5891.5912$\,\AA\ to excite the D$_2$a transition of sodium atoms. An additional 4\,W are emitted in two lines located 1.713\,GHz on either side of the main line at 5891.5714\,\AA\ and 5891.6110\,\AA\ (with 2\,W in each line). Emitting 10\% of the laser power at 5891.5714\,\AA\ is specifically designed to counteract the depopulation of the sodium atoms from the $^2S_{1/2}$ F=2 to F=1 ground state by \textit{re-pumping} them via the D$_2$b transition \citep[where F is the total atomic angular momentum quantum number, see][]{Holzlohner2010}. These three lines are essentially unresolved at the spectral resolution of MUSE, so that in the remainder of this article we simply refer to the 4LGSF laser carrier wavelength in air $\lambda_\text{4LGSF}=5889.959$\,\AA, where:
\begin{equation}
\lambda_\text{4LGSF} =\frac{ \lambda_\text{4LGSF,vacuum}}{n}
\end{equation} 
with $n$ the refractive index of dry air at a pressure of 1 atm and a temperature of 15$^{\circ}$C defined by \cite{Birch1994}:
\begin{equation}
n = 1 + \frac{8.34254}{10^5} + \frac{2.406147}{10^2\times(130-s^2)} + \frac{1.5998}{10^4\times(38.9-s^2)}
\end{equation}
with:
\begin{equation}
s = \frac{10^4}{ \lambda_\text{4LGSF,vacuum}}
\end{equation}

In normal operations, MUSE will employ dedicated $\sim$225\,\AA-wide notch filters -- centered on the LGS frequency -- to block $99.7\pm0.2\%$ of the light emitted by the uplink laser beams. Comparing two back-to-back 60\,sec MUSE exposures (acquired with and without the 4LGSF lasers propagating) readily reveals the presence of a residual narrow \& unresolved emission line at $\lambda_\mathrm{4LGSF}$ (see Fig.~\ref{fig:nominal}). The presence of the notch filter in the light path (leading to the $\sim$225\,\AA\ gap in the spectrum) reduces the intensity of this laser line, without fully blocking it. As such, this line effectively offers an excellent wavelength and spectral resolution reference, given that the laser wavelength is controlled with a root-mean-square error of 0.03\,m\AA. 

We also detect two additional spectral features associated with the use of the 4LGSF at $\lambda\cong6484$\,\AA\ and $\lambda\cong6827$\,\AA. These features consist of 1) a sharp and unresolved emission line with a peak flux density similar to that of the brightest sky emission lines in the 7000\,\AA $\rightarrow$ 8000\,\AA\ spectral range, and 2) broad wings at a level of $< \above 0pt \sim$ $5\times10^{-20}$\,erg\,s$^{-1}$\,cm$^{-2}$\,\AA$^{-1}$ spanning $\sim$130\,\AA\ and $\sim$190\,\AA, respectively. These spectral features evidently result from the inelastic Raman scattering by the O$_2$ and N$_2$ molecules of the laser photons while on their way up to the sodium layer: a process seen easily at high peak powers with e.g. pulsed Light Detection and Ranging (LIDAR) lasers, and exploited for many years in atmospheric physics experiments \citep[see e.g.][]{Leonard1967, Cooney1968, Melfi1969, Collis1970, Melfi1972, Farah2002, Turner2006}. Specifically, laser photons lose through the inelastic collision with O$_2$ or N$_2$ molecules a quantized amount of energy through the excitation of the fundamental $\nu_{1 \leftarrow 0}$ rotational-vibrational (rovibrational) modes of these diatomic molecules \citep{Rasetti1929,Herzberg1950,Whiteman1992}. The nature of Raman scattering is such that the molecule responsible for a specific spectral line is not tied to the absolute wavelength of the line, but instead to its shift from the exciting (laser) wavelength. Indeed, the spectral lines in our MUSE observations are in perfect agreement with the tabulated Raman shifts of $\Delta\nu_{0,\mathrm{O}_2}=1556.4$\,cm$^{-1}$ and $\Delta\nu_{0,\mathrm{N}_2}=2330.7$\, cm$^{-1}$ for O$_2$ and N$_2$ \citep{Herzberg1950}, implying a resulting (post-collision) photon wavelength of $6484.39$\,\AA\ and $6827.17$\,\AA.

\begin{figure}[htb!]
\centerline{\includegraphics[width=\columnwidth]{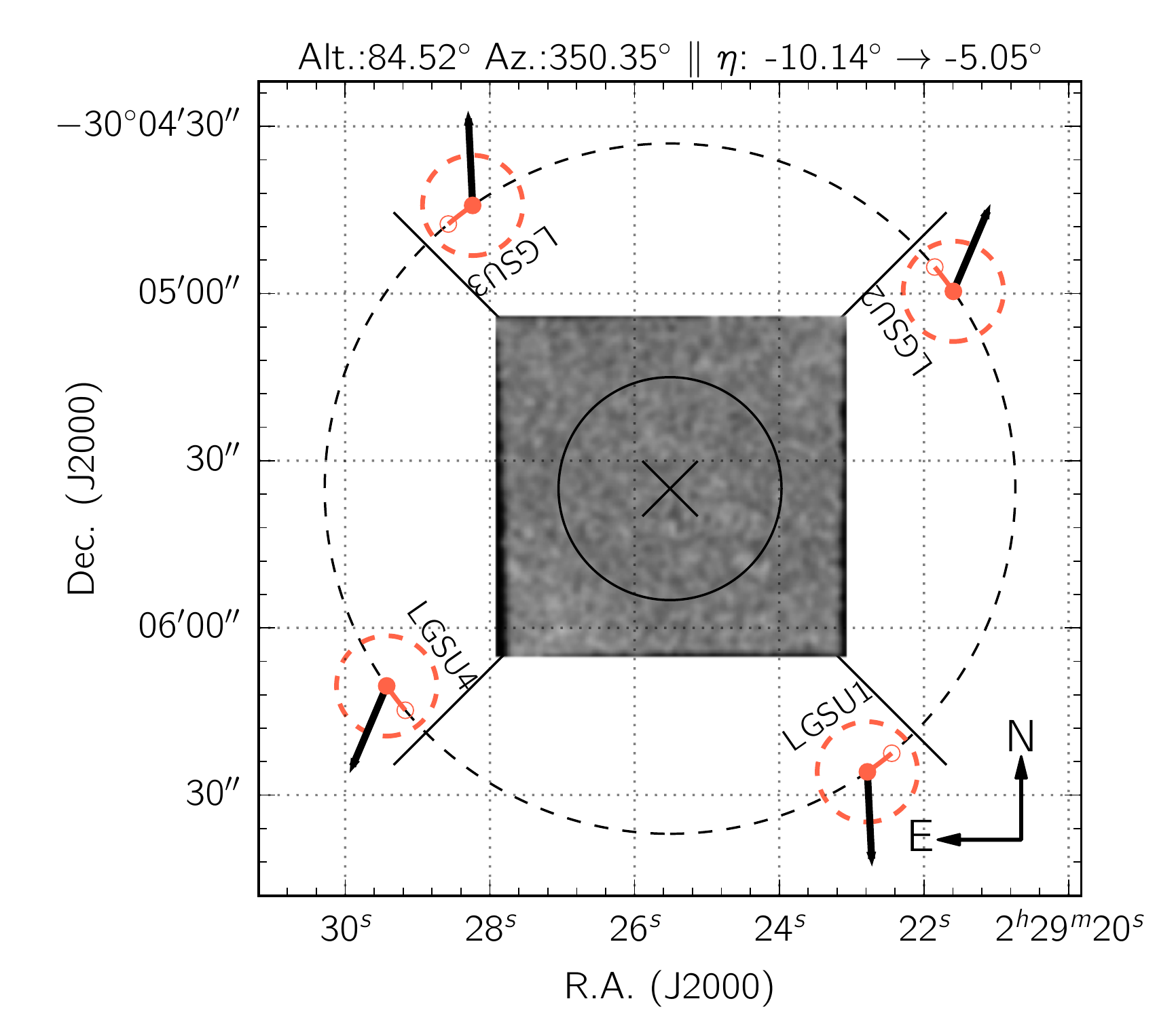}}
\caption{Location of the LGSs (red dots) with respect to the MUSE field in WFM, shown here using the summed spectral slices within 4\,\AA\ from $\lambda_\mathrm{4LGSF}$ and smoothed with a gaussian kernel of 3 pixels in radius from a 60\,s exposure. The LGSs are located 62 arcseconds away from the field center (marked by a black cross) in a fixed altitude-azimuth orientation. The derotation of the MUSE field over time implies that the LGSs rotate around the field center following the parallactic angle $\eta$, with their final position (in this specific case) marked with an empty red dot. The size of the out-of-focus LGS donuts is marked using dashed red circles. Black arrows indicate the orientation of the laser beam from each LGS, pointing away from the MUSE field. As a size reference, the central circle is 20 arcseconds in radius.}\label{fig:LGS_loc}
\end{figure}

\begin{figure}[htb!]
\centerline{\includegraphics[width=\columnwidth]{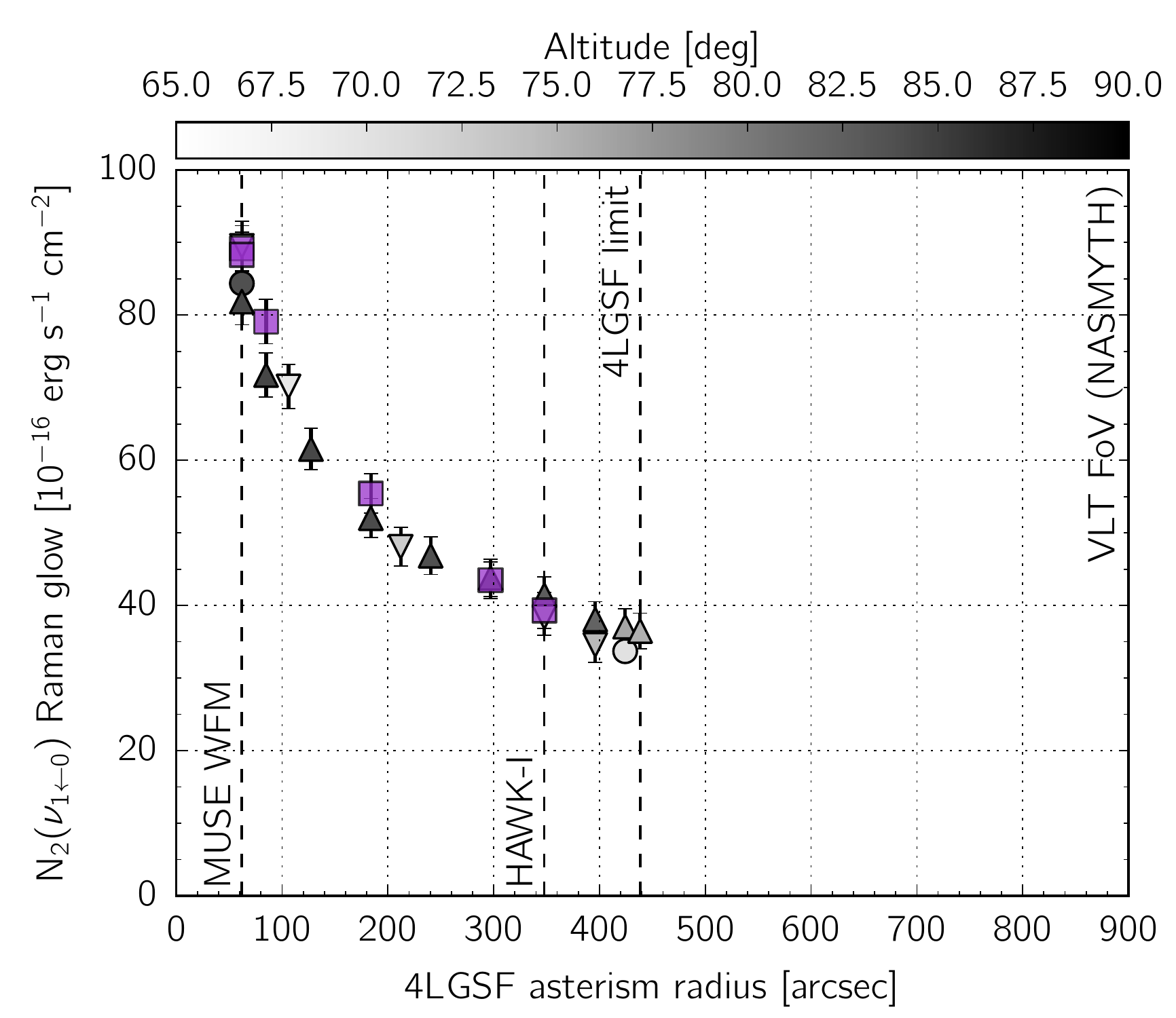}}
\caption{Variation of the intensity of the Raman glow (measured via the flux of the N$_2$($\nu_{1\leftarrow0}$) main vibrational line) within 20 arcseconds of the center of the MUSE WFM field, for different 4LGSF asterism sizes. Different symbols denote distinct series of observations at different azimuths and times. Each grey symbol is color-coded as a function of the telescope altitude at the time of the observation (pre-GALACSI installation). Measurements acquired on 2017, April 15 post-GALACSI installation are shown with purple squares.  Error bars indicate the 3-$\sigma$ uncertainties.}\label{fig:glow}
\end{figure}

In normal operations, the four LGSs are located outside of the MUSE field-of-view ($\sim$62 arcseconds from the center in WFM; see Fig.~\ref{fig:LGS_loc}). As such, the Rayleigh and Raman signals in the MUSE WFM observations presented in Fig.~\ref{fig:nominal} (i.e., with the LGSs outside of the instrument field-of-view) are best described as a residual \textit{glow} throughout the field-of-view. As illustrated in Fig.~\ref{fig:glow}, we note that 1) the intensity of the Rayleigh (and Raman) glow is strongly dependent on the size of the 4LGSF asterism (probed out to its physical limit of $\sim$440 arcseconds in radius set by the opto-mechanical range of the 4LGSU field steering devices), and 2) there is no evidence for the Rayleigh (and Raman) glow to be dependent on the altitude of the observation. Preliminary, follow-up observations performed on 2017, April 15 during the second commissioning run of GALACSI reveal that the installation of this AO module does not have any noticeable influence on the intensity of the Raman lines seen by MUSE. The commissioning of GALACSI and the AO mode of MUSE is ongoing; a detailed and comprehensive characterization of the intensity and impact of Raman lines on MUSE AO observations (both in the WFM and NFM) is thus clearly not feasible just yet. This aspect will be thoroughly investigated during commissioning observations, and reported on separately. 

\begin{figure*}[htb!]
\centerline{\includegraphics[width=\textwidth]{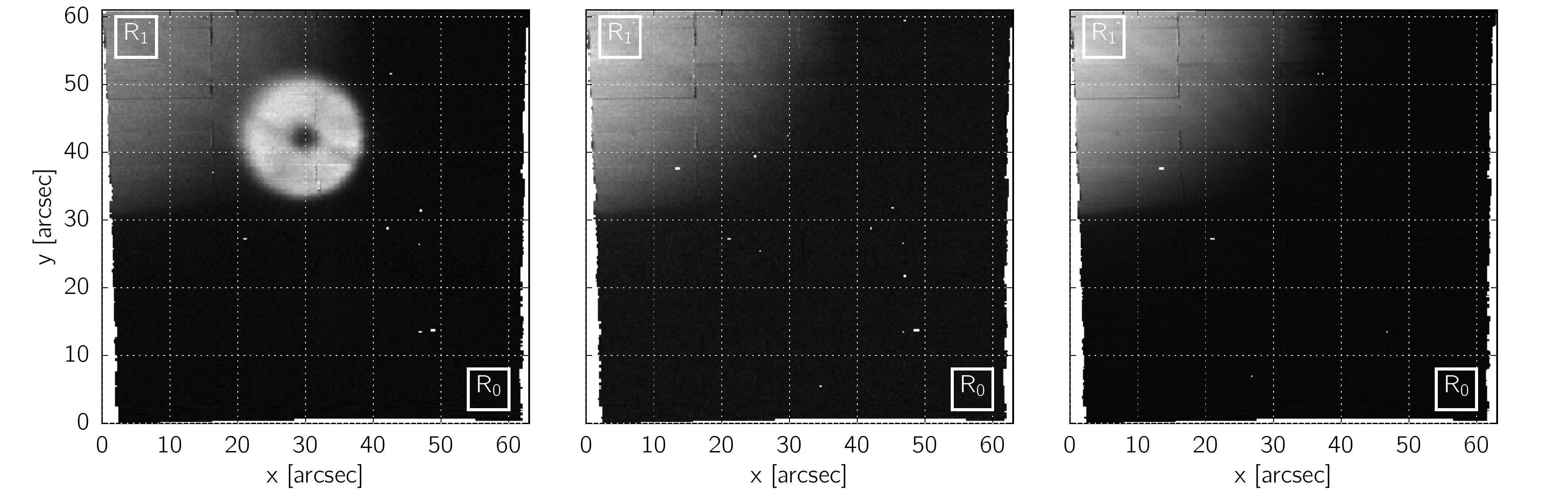}}
\centerline{\includegraphics[width=\textwidth]{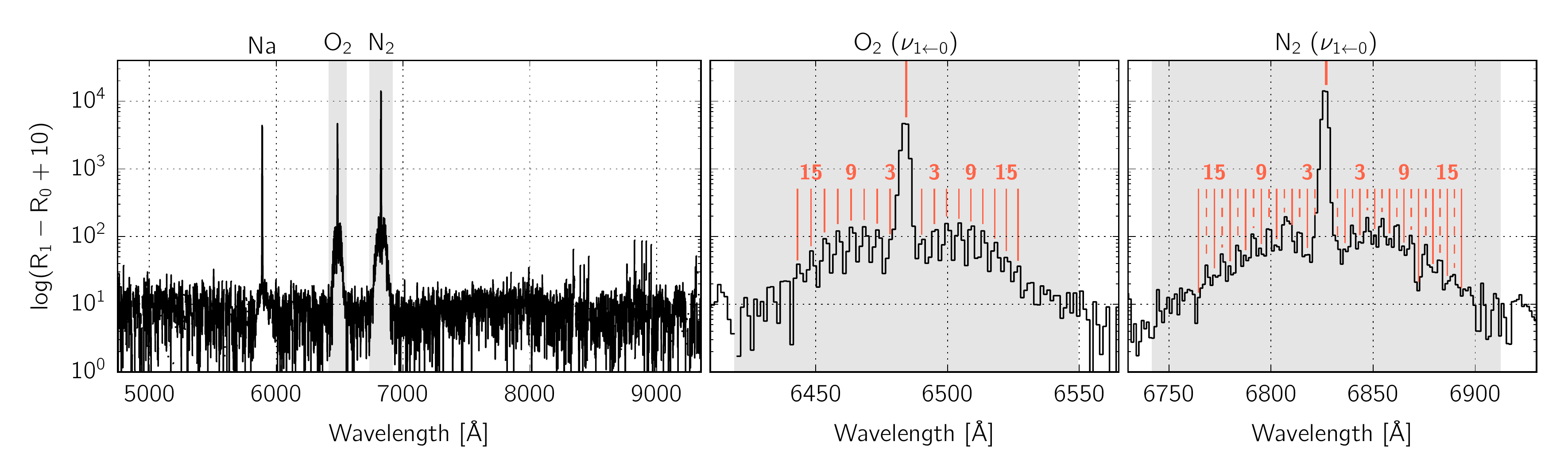}}
\caption{Top: summed MUSE spectral slices within 4\,\AA\ from the laser line at 5889.959\,\AA\ (left), the O$_2$ Raman vibrational line at $6484.39$\,\AA\ (middle), and the N$_2$ Raman vibrational line at $6827.17$\,\AA\ (right), with one of the LGS unit voluntarily placed within the field of view. Both the out-of-focus donut of the LGS (emitted at the sodium layer altitude of $\sim$90\,km), and the laser uplink beam (via Rayleigh and Mie scattering) are visible \textit{through} the notch filter at the laser wavelength. At the Raman wavelengths, only the laser beam is visible. Bottom: difference between the in-beam and out-beam MUSE spectra extracted from the regions R$_1$ and R$_0$ defined in the top panels. Only three features clearly dominate the residual noise: the laser line at $\lambda_\mathrm{4LGSF}$ and the rovibrational Raman lines associated with the O$_2$ and N$_2$ molecules. For O$_2$, red vertical lines show the predicted locations of the rotational side-lines for $J\in[1,3,5,7,\ldots,17]$. For N$_2$, red vertical lines show the predicted locations of the rotational side-lines for $J\in[0,1,2,3,\ldots,17]$: full and dashed for odd and even values of $J$ (respectively).}\label{fig:on-axis}
\end{figure*}

The absence of GALACSI in the light path during the 4LGSF commissioning runs of 2016 June, November \& December did not, however, preclude a detailed characterisation of the Raman emission lines themselves. For test purposes, we acquired a 60\,sec MUSE exposure with one LGS guide star \textit{voluntarily} placed within the $1\times1$ square arcminute field-of-view of MUSE in WFM-N (see Fig.~\ref{fig:on-axis}). At the laser wavelength $\lambda_\mathrm{4LGSF}$, this observation reveals the out-of-focus LGS \& thus the UT4 pupil (in this case prior to the installation of the DSM), as well as the uplink laser beam. At the mean wavelength of the rovibrational Raman lines, the same dataset reveals a clear view of the laser beam, but not of the LGS donut. The LGS is created through the excitation of Na atoms at $\sim$90\,km of altitude, so that the lack of Raman emission associated with the LGS itself is to be expected. We note that the laser beam is not continuously visible (either through Rayleigh or Raman scattering) all the way to the sodium layer. Using a model of the UT4 telescope and 4LGSF launch telescope (see the Appendix~\ref{sec:alt} for details), we find that the Rayleigh and Raman emission from the laser beam is undetected in our observations for altitudes beyond $30\pm5$\,km above ground, consistent with atmospheric LIDAR observations \citep{Keckhut1990,Liu2014}, and typical atmospheric density profiles \footnote{See for example the international Standard Atmosphere model; ISO 2533:1975.}.

Although not representative of normal operations, the on-axis observation of one of the LGS and its associated uplink beam with MUSE allows the construction of a high signal-to-noise view of the Raman rovibrational lines associated with O$_2$ and N$_2$. The mean spectrum of the regions $\mathrm{R}_1-\mathrm{R}_0$ (each 6$\times$6 square arcseconds in size) presented in Fig.~\ref{fig:on-axis} is built to that end. Following \cite{Herzberg1950}, the Raman spectral shifts of the ro-vibrational lines $\Delta\nu_{1\leftarrow0}|_S$ (for the $S$-branch), $\Delta\nu_{1\leftarrow0}|_O$ (for the $O$-branch) and $\Delta\nu_{1\leftarrow0}|_Q$ (for the Q-branch) for a vibrating \& rotating homonuclear diatomic molecule are:
\begin{equation}
\left.\Delta\nu_{1\leftarrow0}\right|_S = \Delta\nu_0+6B_1+(5B_1-B_0)J+(B_1-B_0)J^2, 
\end{equation}
\begin{equation}
\left.\Delta\nu_{1\leftarrow0}\right|_Q = \Delta\nu_0+(B_1-B_0)J+(B_1-B_0)J^2,
\end{equation}
with $J\in[0,1,2,\ldots]$, and:
\begin{equation}
\left.\Delta\nu_{1\leftarrow0}\right|_O = \Delta\nu_0+2B_1-(3B_1+B_0)J+(B_1-B_0)J^2,
\end{equation}
with $J\in[2,3,4,\ldots]$. $B_i$ is the rotational constant of the diatomic molecule at the vibrational level $i$, which can be approximated as:
\begin{equation}
B_i = B_e -\alpha_e(i+0.5),
\end{equation}
with $B_e$ the rotational constant in equilibrium position and $\alpha_e$ the first order term. From \cite{Huber1979,Irikura2007} and references therein: 
\begin{eqnarray}
B_{e,\mathrm{O}_2}=1.44562\mathrm{\,cm}^{-1},\ \alpha_{e,\mathrm{O}_2}=0.0159305\mathrm{\,cm}^{-1},\\ 
B_{e,\mathrm{N}_2}=1.998241\mathrm{\,cm}^{-1},\ \alpha_{e,\mathrm{N}_2}=0.017318\mathrm{\,cm}^{-1}.
\end{eqnarray}
The spectral location of the observed rotational side-bands (visible up to rotational quantum numbers $J=17$ in Fig.~\ref{fig:on-axis}) are in perfect agreement with the theoretical predictions of the Raman shifts of the $S$- and $O$-branch, while the $Q$-branch remains unresolved by MUSE. We also note that for O$_2$, only rotational side-lines for odd values of $J$ are visible, while for N$_2$, rotational side-lines for even values of $J$ are stronger than for odd values of $J$ \citep{Herzberg1950,Barrett1968}.

\section{Extending the search for Raman scattering beyond the $\nu_{1\leftarrow0}$ fundamental mode of O$_2$ and N$_2$ molecules}\label{sec:more}

Virtually all of the Rayleigh, Mie and Raman scattering of laser photons occurs within the footprint of the uplink beams: narrow cylinders of 30\,cm in diameter extending from the laser launch telescopes up the sodium layer at an altitude of $\sim$90\,km, and beyond. As illustrated in Fig.~\ref{fig:on-axis}, looking directly at the uplink beam with MUSE offers an ideal way of \textit{boosting our detection ability} for the Raman scattering signatures of different molecules. Beyond N$_2$ and O$_2$ that make up (by volume) 78.09\% and 20.95\% of air, CO$_2$ (0.04\%), H$_2$O (variable, typically $<$1\%) and CH$_4$ ($\sim$0.0002\%) are the next best candidates for triggering laser-induced Raman scattering events in the sky of Cerro Paranal, given the clean air of the surrounding Atacama desert. 

To further increase the signal of Raman-scattered photons compared to the observations described in Fig.~\ref{fig:on-axis} (with one LGS within the MUSE field-of-view), we acquired an additional series of observations with MUSE in which we \textit{crossed} the laser beams over the field-of-view. One must stress once more that this is a setup clearly not representative of normal science operations. We performed these observations in four individual steps:
\begin{enumerate}
\item setup the 4LGSF asterism to the MUSE WFM size (62 arcseconds in radius),
\item stop the propagation of all lasers but LGSU2, and place its LGS at the location of LGSU4,
\item without modifying the position of LGSU2, start the propagation of LGSU4 and place its LGS at the \textit{original} location of LGSU2, and
\item repeat the same procedure for the LGSUs 1 \& 3.
\end{enumerate}

\begin{figure*}[htb!]
\centerline{\includegraphics[width=\textwidth]{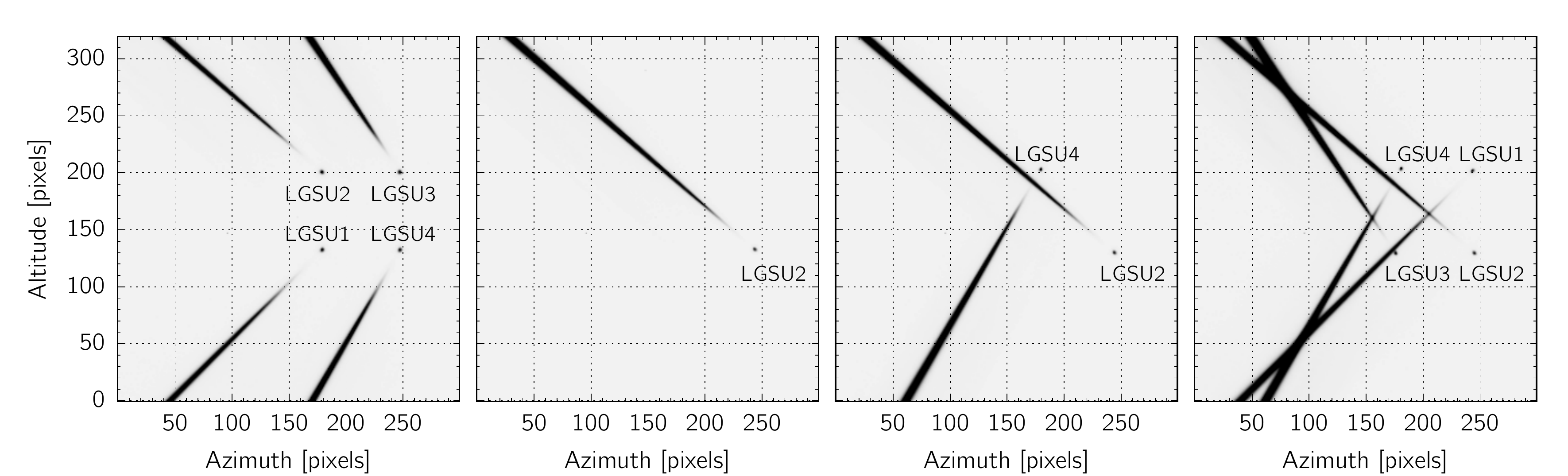}}
\caption{Images from the LPC \citep[1.29 arcseconds per pixel;][]{BonacciniCalia2014a,Centrone2016} with (from left to right): the 4LGSUs in their nominal MUSE-WFM asterism (radius of $\sim$62 arcseconds); LGSU2 at the position of LGSU4; LGSU2 and LGSU4 at each other's default positions; all LGSUs' positions swapped (LGSU1$\rightarrow$ LGSU3, LGSU2$\rightarrow$ LGSU4, LGSU3$\rightarrow$ LGSU1, and LGSU4$\rightarrow$ LGSU2). The off-axis location of the LPC (on the UT4 top ring) gives rise to the peculiar perspective of the \textit{crossed} configurations.}\label{fig:LPC}
\end{figure*}

This process is illustrated in Fig.~\ref{fig:LPC} by the means of images from the Laser Positioning Camera \citep[LPC;][]{BonacciniCalia2014a,Centrone2016} located on the top-ring of UT4. The unusual geometry of the beams in the \textit{crossed} configurations is a direct consequence of the off-axis location of this camera. The lack of a single intersection point for the four laser beams is due to the rectangular distribution of the launch telescopes, compared to the square shape of the LGS asterism. From the MUSE perspective, this implies that the different beams will together excite several altitude layers of the atmosphere below 10\,km within the instrument's WFM field-of-view of 1 square arcminutes.

\begin{figure*}[htb!]
\centerline{\includegraphics[width=\textwidth]{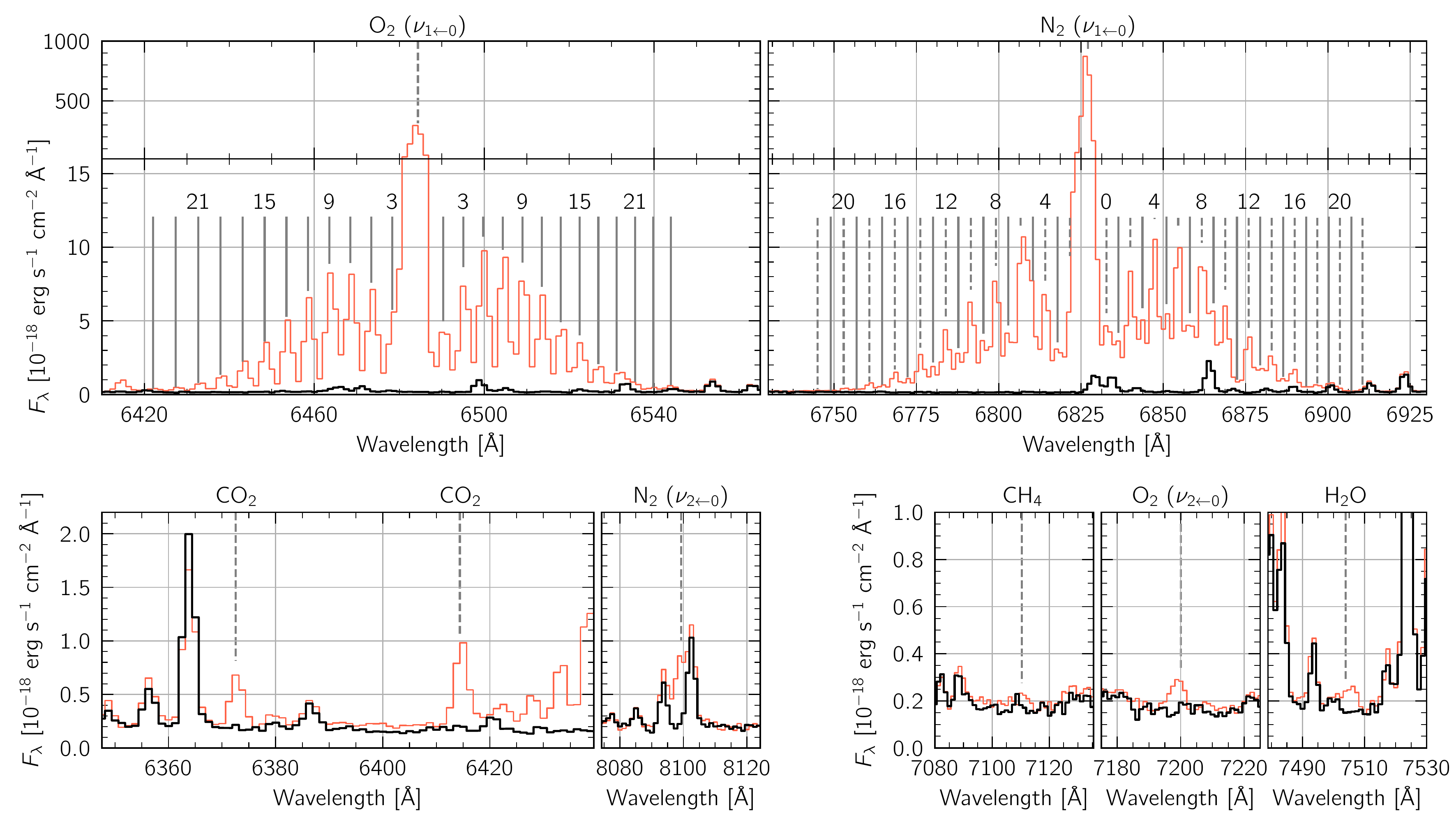}}
\caption{Mean spectra from MUSE spaxels located within 20 arcseconds from the center of the field, with the four LGS beam crossed over the field (red), and with no laser propagating (black). Each panel is centered on a Raman line of interest (which unlike sky lines have no counterpart in the reference spectrum in black). The rotational side-lines associated with the $\nu_{1\leftarrow0}$ transition from the O$_2$ and N$_2$ molecules are detected up to $J=23$. The fundamental Raman lines from CO$_2$ and H$_2$O are clearly visible when the lasers are propagating in this configuration. The first overtone associated with the $\nu_{2\leftarrow0}$ transition from O$_2$ and N$_2$ are also detected (the latter between two sky lines).}\label{fig:H2O}
\end{figure*}

The mean spectra extracted from all the spatial pixels (spaxels) located within 20 arcseconds from the center of the MUSE WFM field-of-view is presented in Fig.~\ref{fig:H2O}, both for a 60\,s exposure with the four laser beams \textit{crossed} and a subsequent reference 60\,s exposure with no laser propagating. The Raman lines associated with the fundamental $\nu_{1\leftarrow0}$ rovibrational excitation of the O$_2$ and N$_2$ molecules are clearly detected, with rotational side-bands visible up to $J=23$ (see Eq.~1 and 2). 

Beyond the fundamental Raman lines, we also detect the first overtone lines \citep[corresponding to the $\nu_{2\leftarrow0}$ transition;][]{Creek1975,Knippers1985,Heaps1996} from O$_2$ and N$_2$ molecules at 7200.02\,\AA\ ($\Delta\nu^{\prime}_{0,\mathrm{O}_2}=3089.2\,\mathrm{cm}^{-1}$) and 8099.23\,\AA\ ($\Delta\nu^{\prime}_{0,\mathrm{N}_2}=4631.2\,\mathrm{cm}^{-1}$). These second order lines are 3-4 orders of magnitude fainter than the peak intensity of the fundamental vibrational line of N$_2$. 

At similar intensity levels, we unambiguously detect the presence of Raman scattering from CO$_2$ molecules at 6372.57\,\AA\ and 6414.39\AA\ \citep[$\Delta\nu_{0,\mathrm{CO}_2}=\left\{1285.8\,\mathrm{cm}^{-1}; 1388.1\,\mathrm{cm}^{-1}\right\}$;][]{Herzberg1945} and H$_2$O molecules at 7503.93\,\AA\ \citep[$\Delta\nu_{0,\mathrm{H}_2\mathrm{O}}=3651.7\,\mathrm{cm}^{-1}$;][]{Herzberg1945,Penney1976}. We note that these observations were acquired with low levels of precipitable water vapor ($\sim$1\,mm) and humidity ($\sim$10\%). The signature of Raman scattering from CH$_4$ molecules at 7110.43\,\AA\ \citep[$\Delta\nu_{0,\text{CH}_4}=2914.2$\,cm$^{-1}$;][]{Herzberg1945} is only tentatively detected with a signal-to-noise S/N$<$3.

\section{Quantifying the intensity of laser-induced Raman lines}\label{sec:fluxes}

Predicting the exact impact of Raman scattering from a LGS system on a given astronomical observation is not straightforward. The number of laser photons that are subject to Raman scattering is proportional both to the power (on-sky) of the laser system employed and the density of molecules within the footprint of the uplink beam, which varies as a function of the altitude. In addition, the intensity of the contamination for any given telescope at any given time also depends on the \emph{collision parameters} (i.e. the scattering angle and the distance to the impacting portion of the uplink laser beam from the telescope). Finally, the intensity of the Raman contamination also depends heavily on the characteristics of the instrument involved: for example, the central wavelength and width of the filter for an imager, and the spectral range and resolution for a spectrograph. The specificities of the telescope (incl. the overall throughput and aperture) also play a role. Altogether, this implies that it is very difficult to summarize with precision the impact of laser-induced Raman scattering for all possible instruments, telescopes, LGS systems and observatories. 

\begin{figure*}[htb!]
\centerline{\includegraphics[width=\textwidth]{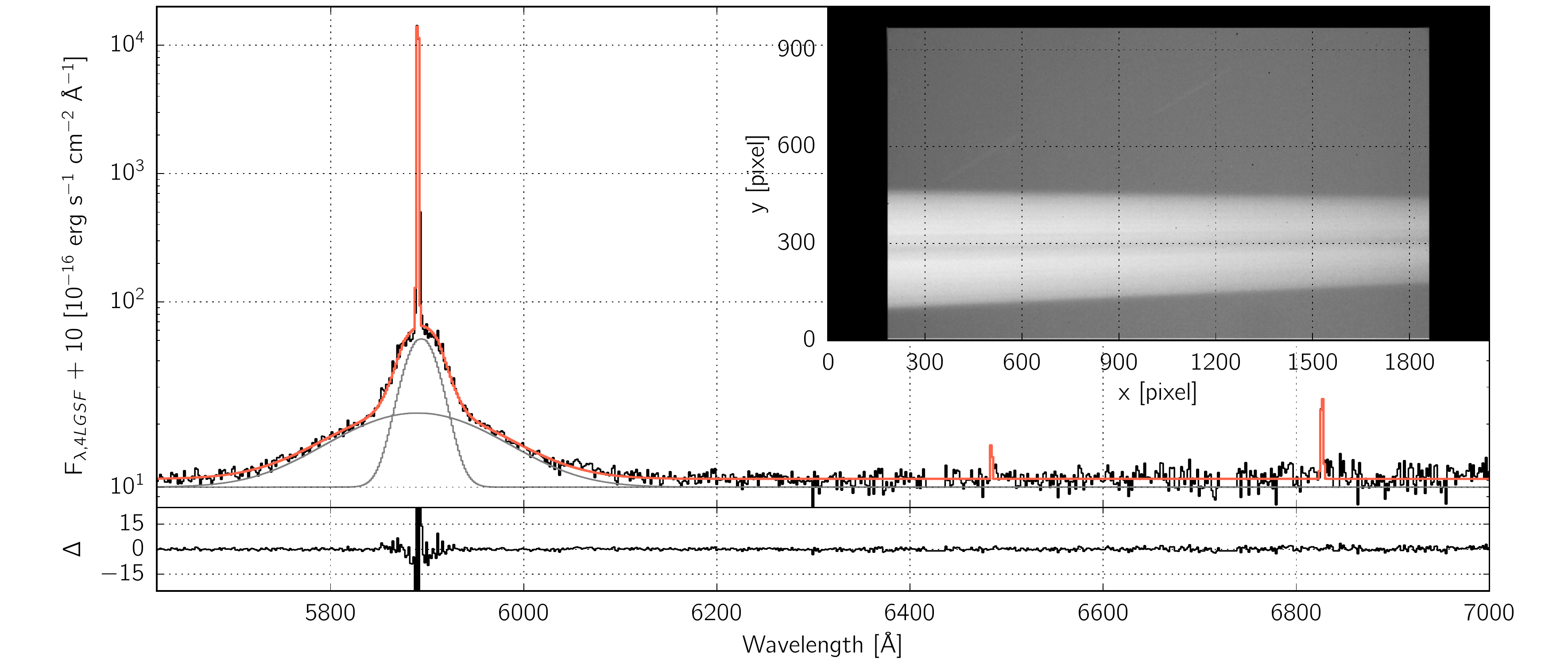}}
\caption{FORS2 spectra of the uplink beam of LGSU1 from the 4LGSF at an altitude of 25\,km above ground (black), and our fit to the spectra (red) including the laser line and the N$_2$($\nu_{1\leftarrow0}$) and O$_2$($\nu_{1\leftarrow0}$) Raman lines. The main laser line at $\lambda_\text{4LGSF}$ requires three Gaussian components to be properly reproduced. The narrowest component corresponds to the laser photons subject to elastic Rayleigh \& Mie scattering. The intermediate Gaussian component corresponds to inelastic rotational Raman scattering, and is in fact a superposition of multiple narrow lines associated with different molecules. The broadest component is an instrumental artefact of FORS2. Top right: R-band image from the chip 1 of FORS2 acquired at the same location, with the uplink beam (horizontal on the image) clearly visible. The \emph{running shadow} of the secondary mirror of UT1 makes the inner-most region of the laser beam darker. Star trails are visible in the background, given that UT1 was idle during the exposure. }\label{fig:fors}
\end{figure*}

\begin{table*}[htb!]
\caption{Integrated fluxes for the different laser-induced Raman lines from the uplink beam at an altitude of 27.6\,km above sea level, normalized to the power of a (continuous wave) 1\,W laser shining at 5889.959\,\AA. Uncertainties on the quoted fluxes are $\lesssim$15\%. For the fundamental Raman lines of N$_2$ and O$_2$, the fluxes refer to the central peak only (i.e. the $Q-branch$). The flux associated with the main Rayleigh line of the laser in the same observational setup is also indicated. \\}\label{tbl:fluxes}
\begin{tabular}{l l c c c c c c c c c}
\hline
\hline
Raman line & & $\lambda_\text{4LGSF}$ & CO$_2$ & CO$_2$ & O$_2$($\nu_{1\leftarrow0}$) & N$_2$($\nu_{1\leftarrow0}$) & CH$_4$ & O$_2$($\nu_{2\leftarrow0}$)  &  H$_2$O &N$_2$($\nu_{2\leftarrow0}$) \\[2pt]
\hline
Raman shift $\left[\text{cm}^{-1}\right]$ & &{--} &{1285.8} &{1388.1} &{1556.4} & {2330.7}  &{2914.2} & {3089.2} & {3651.7} &{4631.2}\\[2pt]
$\lambda_{obs}$ $\left[\text{\AA}\right]$ & & {5889.959} &{6372.57} &{6414.39} &{6484.39} & {6827.17} &{7110.43} &{7200.02} &{7503.93} &{8099.23}\\[2pt]
Flux $\left[\text{10$^{-20}$\,erg\,s$^{-1}$\,cm$^{-2}$ W$^{-1}$}\right]$& & {$1.9\cdot10^{7}$} &{11.3} &{18.9} &{$6.8\cdot10^{3}$} & {$2.0\cdot10^{4}$} & {$\lesssim1.1$} & {3.2} & {2.7} & {16.1}\\[2pt]
\hline
\end{tabular}
\end{table*}

These factors notwithstanding, we provide in Table~\ref{tbl:fluxes} indicative fluxes of the different Raman lines identified in Sec.~\ref{sec:more}. These fluxes are scaled to the brightness of the uplink beam of a (continuous wave) 1\,W laser guide star with $\lambda_\text{LGS}=\lambda_\text{4LGSF}$ observed at an altitude of 25\,km above ground on Cerro Paranal, equivalent to 27.6\,km above sea level. The appropriate scaling for the different line fluxes (measured from the spectrum presented in Fig.~\ref{fig:H2O}) was derived using dedicated observations of one of the 4LGSF uplink beam using the FORS2 \citep{Appenzeller1998} spectrograph mounted on the Cassegrain focus of UT1 at the VLT, performed during the evening twilight of 2016, November 28 under clear conditions. During the test, UT4 was pointed directly \emph{away} from UT1 with LGSU1 (alone) propagating in center-field at an altitude of 89.000$^{\circ}$. UT1, located at a (Vincenty) distance of 130.22\,m from UT4, was pointed directly \emph{towards} UT4 at an altitude of 88.694$^{\circ}$. In this configuration (with both telescopes idle during the entire duration of the test), UT1 was effectively pointing towards the uplink laser beam at an altitude of 25\,km above ground.

The resulting FORS2 long-slit spectrum for a 60\,s exposure acquired with a slit-width of 0.28 arcseconds and the 600RI+19 grism is presented in Fig.~\ref{fig:fors}. The data were processed using the ESO Reflex \citep{Freudling2013} workflow (v 5.0.20) for FORS2. We use Gaussian components to fit the main laser line and the fundamental Raman lines associated with O$_2$ and N$_2$ molecules, combined with a constant level to account for the existing continuum residual post sky subtraction. We note that fitting the main laser line at $\lambda_\text{4LGSF}$ requires 3 Gaussian components. The narrowest (unresolved) component is the brightest, and corresponds to the laser photons that experienced elastic Rayleigh and Mie scattering. The intermediate component, with a full-width-at-half-maximum FWHM = 44.2$\pm$0.6\,\AA, corresponds to the laser photons that experienced Raman scattering and excited the \emph{rotational} mode of molecules (only, as opposed to the \emph{rovibrational} modes described until now). This component is as such a collection of narrow emission lines --blended at the spectral resolution of FORS2 of R$\cong$2900-- associated with the different air molecules present in the footprint of the laser beam. The existence of (some of) these narrow emission lines is best seen in the fit residual $\Delta$. Finally, the broadest Gaussian component associated with the main laser line is an instrumental artefact of the FORS2 spectrograph. A similar glow is present, for example, around the brightest lines nearing saturation in arc lamp exposures. The exact origin of this glow is under investigation by the FORS2 instrument team.  

We derive the appropriate scaling factor for all the Raman line fluxes presented in Table~\ref{tbl:fluxes} by comparing the FORS2 observations of the N$_2$ and O$_2$ fundamental Raman lines with the MUSE observations described in Sec.~\ref{sec:more}. The resulting fluxes thus correspond to the integrated flux of a single laser uplink beam observed at an altitude of 25\,km, under the (reasonable) assumption that the composition of the atmosphere above Cerro Paranal does not vary significantly up to heights of $\sim$25\,km above ground (i.e. the intensity ratio between the different laser-induced Raman lines is constant).

The line fluxes provided in Table~\ref{tbl:fluxes} are intended for observatories and their users to assess the possible impact of laser-induced Raman scattering on regular operations. We stress however that these fluxes must be treated with caution. They provide a realistic measure of the different Raman line fluxes, but their validity is limited to the meteorological specificities of a given site and seasonal variations in the temperature profile and water vapor content of the atmosphere. For reference and comparison purposes, the IR temperature of the sky was ($-88\pm5$)$^{\circ}$C during the FORS2 observations, with a temperature at ground level  of ($16\pm1$)$^{\circ}$C. For the H$_2$O line in particular, the precipitable water vapor during the (scaled) MUSE observations was $\sim$1\,mm and the humidity $\sim$10\% at ground level.

\section{Implications of laser-induced Raman scattering for astrophysical observations}\label{sec:astro}
\begin{figure}[htb!]
\centerline{\includegraphics[scale=0.5]{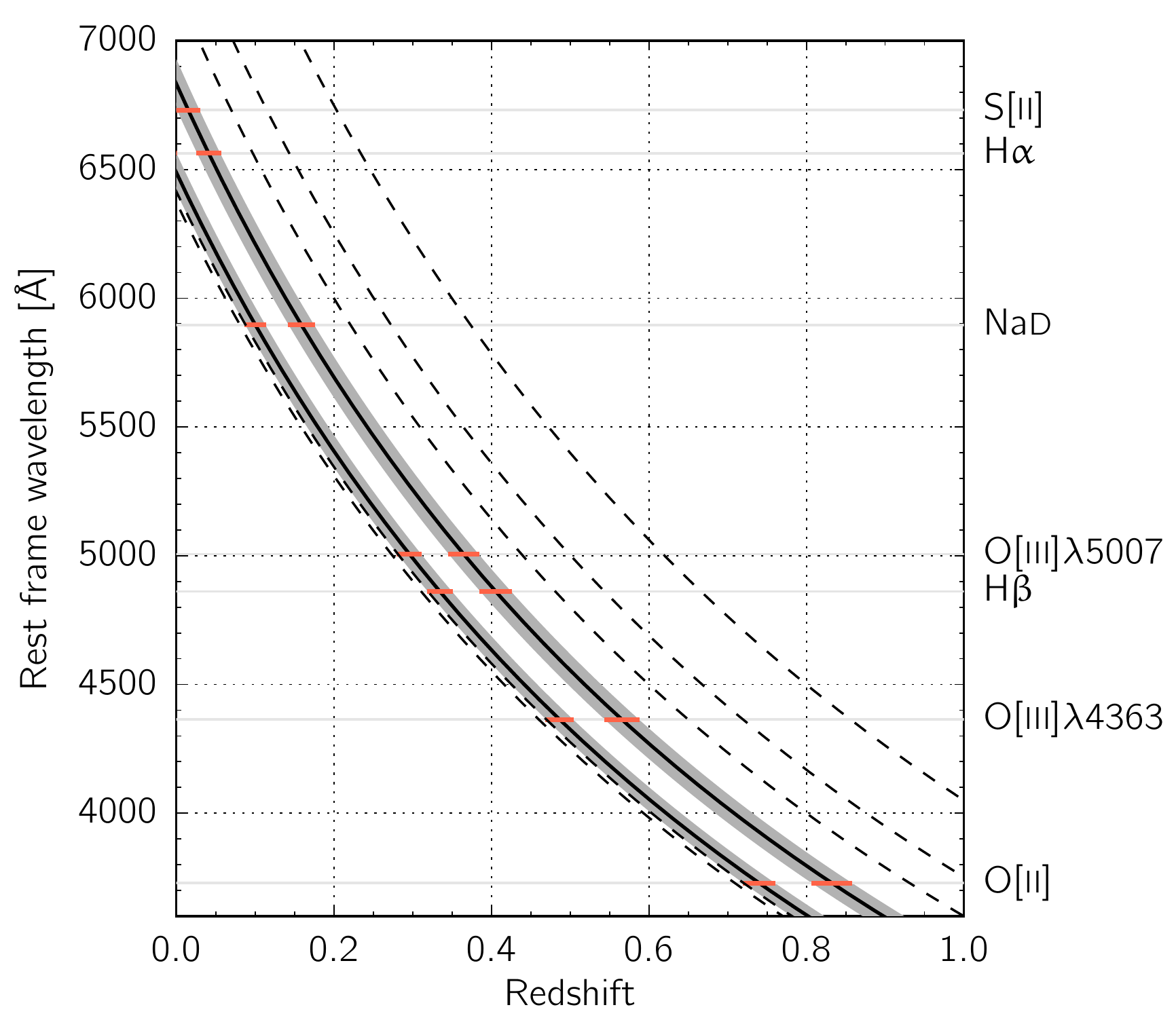}}
\caption{Spectral region -- in the rest-frame of an astrophysical source -- affected by laser-induced Raman contamination as a function of redshift, assuming $\lambda_\text{LGS}=589$\,nm. For both O$_2$ and N$_2$ molecules, the greyed-out zone marks the spectral extent of the rotational side-lines out to $J=17$. The redshift zones associated with a few important optical lines are highlighted in red. Spectral regions affected by the (three to four orders of magnitude fainter) Raman contamination from CO$_2$ molecules, H$_2$O molecules, and the first overtone of O$_2$ molecules and N$_2$ molecules are indicated with dashed-dotted lines.}\label{fig:redshift}
\end{figure}

Elegant from the perspective of the physics involved, the possible contamination of observations via laser-induced rovibrational Raman scattering may become a clear concern from the operational perspective of an astronomical observatory. As illustrated in Fig.~\ref{fig:on-axis}, rovibrational Raman scattering leads to spectrally complex and broad structures in optical regions containing important astrophysical spectral lines (see Fig.~\ref{fig:redshift}). Although the structure of the Raman lines associated with diatomic molecules like N$_2$ and O$_2$ is well known, removing these lines from the data \textit{a posteriori} will necessarily induce some residual photon noise affecting the quality of observations, in particular for deep fields and/or extremely faint targets. The intensity of the Raman rovibrational lines is also temperature-dependent \citep{Cooney1972,Keckhut1990}, and thus potentially time-dependent. The use of notch filters, such as in the case of MUSE, is only viable (from a scientific efficiency perspective) to alleviate the impact of the main Rayleigh line and \emph{rotational} Raman lines in its immediate surroundings. It is thus crucial to handle the impact of rovibrational Raman contamination by first ensuring the proper baffling of the instrument, in addition to designing appropriate calibrations, observation procedures, and data reduction pipelines.

Certainly, the existence of laser-induced Raman scattering is not unique to Cerro Paranal, but common to any astronomical observatory employing similar devices. The majority (today) of astronomical telescopes rely on 589\,nm lasers, but the emergence of so-called \emph{Rayleigh} LGS systems with shorter lasing wavelengths implies that at specific sites, laser-induced Raman scattering will affect different spectral regions. For completeness, we list in Table~\ref{tbl:lams} the wavelengths of the atmospheric rovibrational Raman lines described in Sec.~\ref{sec:more} as a function of the laser wavelength $\lambda_\text{LGS}$ for the main astronomical telescopes equipped with an LGS system. The efficiency of Raman scattering increases at shorter wavelengths, so that \emph{Rayleigh} LGS systems are likely to be more strongly affected by it, modulo the on-sky power the lasers involved.

\begin{table*}[htb!]
\caption{Wavelength of the main atmospheric, laser-induced Raman scattering lines as a function of the LGS wavelength $\lambda_\text{LGS}$ at different astronomical telescopes.\\}\label{tbl:lams}
\begin{tabular}{l l c c c c c c c c c}
\hline
\hline
Telescope & LGS system & $\lambda_\text{LGS}$& CO$_2$ & CO$_2$ & O$_2$($\nu_{1\leftarrow0}$) & N$_2$($\nu_{1\leftarrow0}$) & CH$_4$ & O$_2$($\nu_{2\leftarrow0}$)  &  H$_2$O &N$_2$($\nu_{2\leftarrow0}$) \\[2pt]
 & & [nm] &[nm]&[nm]&[nm]&[nm]&[nm]&[nm]&[nm]&[nm]\\[2pt]

\hline
Gemini North & ALTAIR \citep{Boccas2006,Christou2010}  & \tikzmark{a} &  \tikzmark{c} & \tikzmark{e} & \tikzmark{g} &  \tikzmark{i}& \tikzmark{k} & \tikzmark{m} & \tikzmark{o}& \tikzmark{q}\\[2pt]
Gemini South & GeMS \citep{dOrgeville2012,Rigaut2014,Neichel2014}  &  & & & & & &&&\\[2pt]
W.~M. Keck & LGSAO \citep{Wizinowich2006,vanDam2006} & & & & & &&&\\[2pt]
Lijang 1.8-m Telescope & LGS-AOS \citep{Wei2016} &&&&&&&&\\[2pt] 
Shane 3-m  Telescope& ShaneAO \citep{Gavel2014,Gavel2016} &&&&&&&&\\[2pt]
Subaru & AO188 \citep{Hayano2010,Minowa2012}  &  & & & & & &&&\\[2pt]
VLT (UT4) & 4LGSF \citep{Arsenault2013,BonacciniCalia2014} & &&&&&&&&\\[2pt]
VLT (UT4) & PARLA \citep{Lewis2014} & &&&&&&&&\\[2pt]
\emph{Thirty Meter Telescope} & LGSF \citep{Boyer2014,Boyer2016} &  & & & & & &&&\\[2pt]
\emph{Giant Magellan Telescope} & GMT AO \citep{Bouchez2014}&  & & & & & &&&\\[2pt]
\emph{Extremely Large Telescope} & ATLAS \& MAORY \citep{Diolaiti2010,Thatte2014,Davies2016,Neichel2016} & \tikzmark{b}  & \tikzmark{d} &\tikzmark{f} &\tikzmark{h} &\tikzmark{j} & \tikzmark{l}&\tikzmark{n}&\tikzmark{p}&\tikzmark{r}\\[2pt]

Large Binocular Telescope & ARGOS \citep{Rabien2010,OrbandeXivry2016}& 532 & 571 & 574 & 580 & 607 & 630 & 637 & 660 & 706\\[2pt]
William Hershel Telescope & GLAS \citep{Benn2008}& 515 & 552 & 555 & 560 & 585 & 606 & 612 & 634 & 676 \\[2pt]
Hale 5-m Telescope & PULSE  \citep{Baranec2014} & \tikzmark{s} &  \tikzmark{u} &  \tikzmark{w} &  \tikzmark{y} &  \tikzmark{aa} &  \tikzmark{ac} &  \tikzmark{ae} &  \tikzmark{ag} &  \tikzmark{ai} \\[2pt]
Southern Astrophysical& \multirow{2}{*}{SAM  \citep{Tokovinin2010}}&\multirow{2}{*}{\tikzmark{t}} & \multirow{2}{*}{ \tikzmark{v} } & \multirow{2}{*}{\tikzmark{x}} & \multirow{2}{*}{\tikzmark{z}} &\multirow{2}{*}{\tikzmark{ab}} & \multirow{2}{*}{\tikzmark{ad}} & \multirow{2}{*}{\tikzmark{af}} & \multirow{2}{*}{\tikzmark{ah}} & \multirow{2}{*}{\tikzmark{aj}} \\[-2pt]
\quad Research Telescope &  &  & & & & & & & & \\[2pt]
\hline
\end{tabular}

\tikz[remember picture, overlay]
{\draw ([yshift=1ex]a) -- (b) node[midway, fill=white] {589};
\draw ([yshift=1ex]c) -- (d) node[midway, fill=white] {637};
\draw ([yshift=1ex]e) -- (f) node[midway, fill=white] {641};
\draw ([yshift=1ex]g) -- (h) node[midway, fill=white] {648};
\draw ([yshift=1ex]i) -- (j) node[midway, fill=white] {683};
\draw ([yshift=1ex]k) -- (l) node[midway, fill=white] {711};
\draw ([yshift=1ex]m) -- (n) node[midway, fill=white] {720};
\draw ([yshift=1ex]o) -- (p) node[midway, fill=white] {750};
\draw ([yshift=1ex]q) -- (r) node[midway, fill=white] {810};
\draw ([yshift=1ex]s) -- (t) node[midway, fill=white] {355};
\draw ([yshift=1ex]u) -- (v) node[midway, fill=white] {372};
\draw ([yshift=1ex]w) -- (x) node[midway, fill=white] {373};
\draw ([yshift=1ex]y) -- (z) node[midway, fill=white] {376};
\draw ([yshift=1ex]aa) -- (ab) node[midway, fill=white] {387};
\draw ([yshift=1ex]ac) -- (ad) node[midway, fill=white] {396};
\draw ([yshift=1ex]ae) -- (af) node[midway, fill=white] {399};
\draw ([yshift=1ex]ag) -- (ah) node[midway, fill=white] {408};
\draw ([yshift=1ex]ai) -- (aj) node[midway, fill=white] {425};
}

\end{table*}

In most cases, LGS systems are used to feed (near-)IR instruments, oblivious to the existence of Raman-scattered photons by the O$_2$ and N$_2$ molecules (which are the most abundant in the atmosphere and thus give rise to the strongest signal). Yet, the presence of laser-induced Raman-scattered photons can also affect other near-by telescopes and instruments at sites hosting multiple telescopes, such as Mauna Kea or Cerro Paranal. The distance between the different telescopes is a key factor influencing the frequency of so-called \emph{laser collisions}: for the case of Cerro Paranal, the probability of at least one collision between the uplink laser beams from UT4 and any of the UTs and the VLT Survey Telescope \citep[VST;][]{Arnaboldi1998,Kuijken2011} is estimated to be $>$90\% on any given night \citep[][]{Amico2010}. To avoid the accidental contamination of \emph{laser-sensitive} observations at near-by telescopes, a laser collision prediction tool was initially developed for the W.M. Keck LGS system \citep{Summers2003}, and eventually ported to the observatory of La Palma \citep{Summers2006} and Cerro Paranal \citep[][]{Amico2015}. One essential component of this tools is that the different telescopes on any given site must declare themselves as being \emph{laser sensitive}, or not. On Cerro Paranal, the sensitivity of the different instruments mounted on all the UTs was assessed individually for each system, filter set, grating and grism, depending whether a given setup can \emph{see} (or not) the main laser Rayleigh line at 589\,nm. In other words, the system has so far been oblivious to the possibility of rovibrational Raman contamination. In particular, laser collisions for instrumental setups sensitive to the rovibrational Raman lines only (but not the main laser line) do not trigger any red flag.  As discussed in Sect.~\ref{sec:fluxes}, each instrument will be affected by rovibrational Raman contamination at a different level. Broad-band imagers can be expected to be less impacted than high-resolution spectrographs, and each case must undoubtedly be assessed individually. At the time of publication of this article, the individual \emph{Raman sensitivity} assessment for all affected instruments on Cerro Paranal is underway.

With respect to the next generation of extremely large telescopes, our observations of laser-induced Raman scattering on Cerro Paranal will be particularly applicable to the HARMONI \citep{Thatte2014} visible and near-IR integral field spectrograph to be installed as a first-light instrument on the ELT at Cerro Armazones. HARMONI will be fed corrected wavefronts by the ELT ATLAS LGS system \citep{Fusco2010} that relies on six 589\,nm lasers similar to those of the 4LGSF (according to the current design). HSIM \citep{Zieleniewski2015} is a dedicated software designed to simulate observations with HARMONI. This tool thus likely offers the ideal means for 1) evaluating the impact of laser-induced Raman scattering on HARMONI observations, and 2) implementing suitable observing strategies. 

Finally, the 4LGSF lasers and associated laser-induced Raman emission are currently under (ongoing) close scrutiny, including via dedicated observations using the VST, FLAMES \citep[][]{Pasquini2002} on UT2, as well as with the Gran Telescopio Canarias using the Wendelstein Laser Guide Star \citep{BonacciniCalia2012}. Combined with existing and upcoming MUSE \& GALACSI tests, these datasets are intended to assess directly the impact of laser-induced Raman scattering on specific instruments on Cerro Paranal and at the Observatorio del Roque de los Muchachos.

\begin{acknowledgments}
We thank the AOF Builders \citep{Arsenault2013} for the construction and installation of the AOF on UT4 at the VLT. We are grateful to Alain Smette for stimulating discussions over the course of the commissioning tests, to George Hau, Julien Girard, Thomas Pfrommer, Steffen Mieske, Andreas Kaufer, Tim de Zeeuw and Roland Bacon for their feedback on this article, to Claudia Reyes Saez, Claudia Cid, Susana Cerda and Diego Parraguez for their operational support, and to Joe Anderson for his input regarding the behavior of the FORS2 spectrograph. We also thank the two anonymous referee for their constructive reviews. This research has made use of the following \textsc{python} packages: \textsc{matplotlib} \citep{Hunter2007}, \textsc{astropy}, a community-developed core Python package for Astronomy \citep{AstropyCollaboration2013}, \textsc{aplpy}, an open-source plotting package for \textsc{python} \citep{Robitaille2012}, and \textsc{mpfit}, a Python script that uses the Levenberg-Marquardt technique \citep{More1978} to solve least-squares problems, based on an original \textsc{fortran} code that is part of the \textsc{minpack}-1 package. This research has also made use of \textsc{montage}, funded by the National Science Foundation under Grant Number ACI-1440620 and previously funded by the National Aeronautics and Space Administration's Earth Science Technology Office, Computation Technologies Project, under Cooperative Agreement Number NCC5-626 between NASA and the California Institute of Technology, of the \textsc{aladin} interactive sky atlas \citep{Bonnarel2000}, of \textsc{saoimage ds9} \citep{Joye2003} developed by Smithsonian Astrophysical Observatory, and of NASA's Astrophysics Data System. Based on observations made with ESO Telescopes at the La Silla Paranal Observatory. All the observations described in this article are freely available online from the ESO Data Archive.
\end{acknowledgments}

\appendix*
\section{Constraining the altitude of the Raman emission with a toy model of the UT4 and 4LGSF systems}\label{sec:alt}

\begin{figure*}[htb!]
\includegraphics[width=0.8\textwidth]{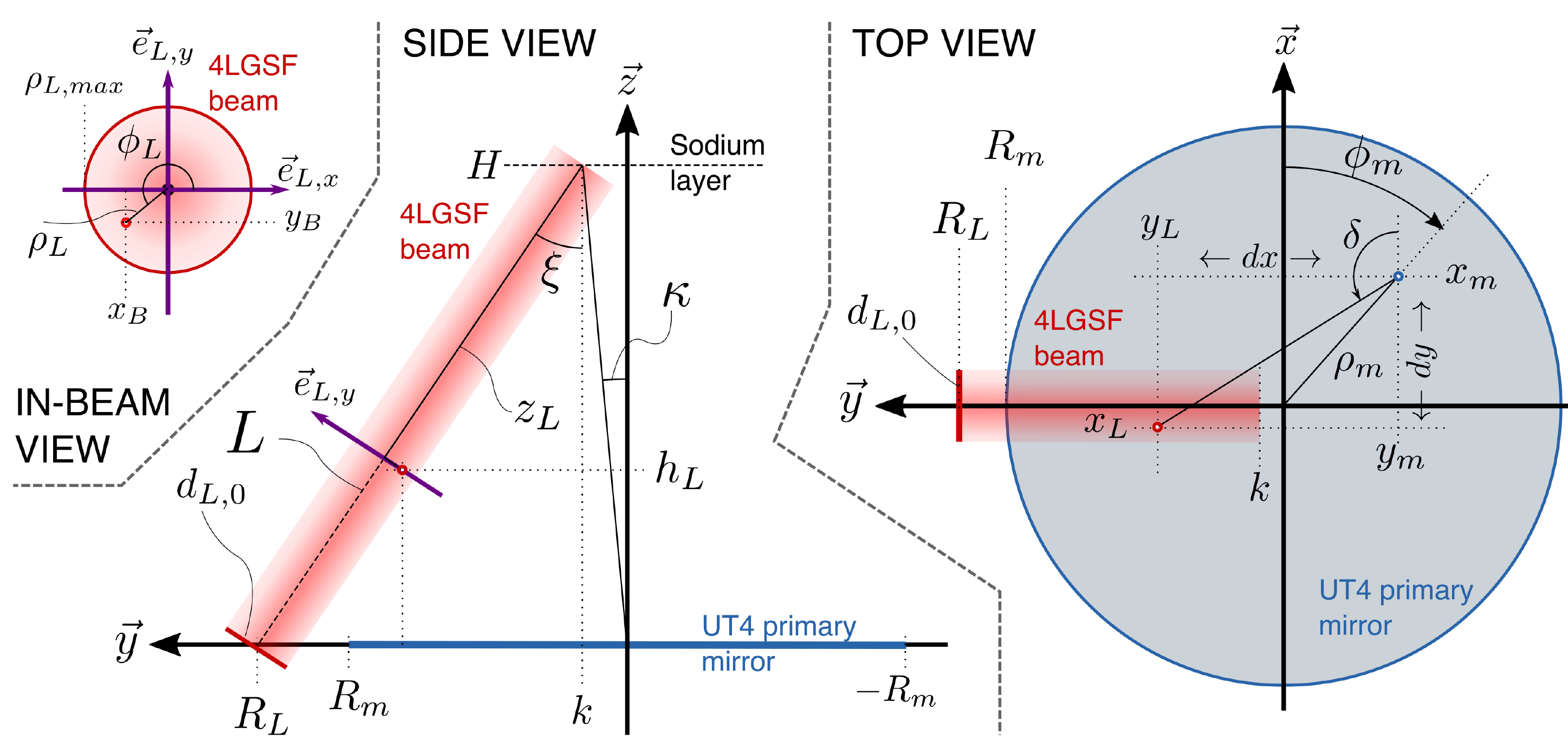}
\caption{Simple model of the 4LGSF (one laser only) and UT4 (reduced to a flat primary mirror). The laser is set to propagate from a distance $R_L$ from the optical axis of the telescope along the $\vec{y}$ axis. The beam is 30\,cm in diameter, and assumed to maintain a constant width as a function of altitude.}\label{fig:schema}
\end{figure*}

The geometry and spatial extent of the Rayleigh emission from the laser beam visible in Fig.~\ref{fig:on-axis} is directly related to the altitude at which the scattering of the laser photons occurs (either via Rayleigh or Raman processes). The geometry of the problem is such that a simple model of the UT4 telescope and 4LGSF launch system can be used to constrain the altitude at which the laser beam is visible within the MUSE field-of-view. A schematic of the 4LGSF systems and UT4 is presented in Fig.~\ref{fig:schema}, along with all the relevant quantities of our model.

We define $R_m=4.1$\,m, the radius of the primary mirror of UT4. Let us further assume that:
\begin{eqnarray}
H &=& 90000\mathrm{\ m}\\
d_{L,0} &=& 0.3\mathrm{\ m} \\
R_L &=& 5.51\mathrm{\ m} 
\end{eqnarray}
where $H$ is the altitude of the 4LGSF guide star (i.e. the altitude of the sodium layer above the primary mirror), $d_{L,0}$ is the diameter of the laser beam, and $R_L$ is the distance between the optical axis of UT4 and the location where the 4LGSF laser beam would cross the mirror plane (if extended backward from the launch telescope). We also assume that the laser beam (which is effectively a gaussian beam) has a constant diameter as a function of the altitude, and that the observation is performed at zenith.

We define $\kappa$ as the angle between the optical axis of UT4 and the location of the 4LGSF guide star, as measured from the center of the primary mirror. This angle corresponds to a physical offset projected onto the mirror surface $k$:
\begin{equation}
k = H\tan{\kappa}
\end{equation}
When the laser guide star is placed at the center of the MUSE field-of-view, $\kappa=0\Rightarrow k=0$. The total path length of the laser beam to the sodium layer $L$ is:
\begin{equation}
L = \sqrt{(R_L-k)^2+H^2},
\end{equation}

To reconstruct the footprint of the laser beam in the MUSE field-of-view, we perform a crude \textit{ray-tracing} exercise and construct the observed position (in the MUSE field-of-view) of all sky locations illuminated by the 4LGSF beam. Each location within the laser beam can be described with $z_L$, its distance to the sodium layer (along the beam propagation axis), $\rho_L$ its distance from the beam center, and $\phi_L$ its azimuth. In the subsequent analysis, we sample the laser beam at intervals of 1\,km along its propagation axis, and each slice with 3 radial and 10 azimuthal steps. 
For a chunk of atmosphere illuminated by the laser with coordinates ($z_L$;$\rho_L$;$\phi_L$), its projected coordinates in the mirror plane ($x_L$;$y_L$) are:
\begin{eqnarray}
x_L &=& x_B\\
y_L &=& y_B\cos\xi + z_L\sin\xi + k
\end{eqnarray}
with:
\begin{eqnarray}
x_B &=& \rho_L\cos\phi_L\\
y_B &=& \rho_L\sin\phi_L
\end{eqnarray}
Here, we assume a geometry in which the laser beam is launched from the $\vec{y}$-axis at the location (0;$R_L$), with the laser $\vec{e}_{L,x}$-axis in the $\vec{x}-\vec{z}$ plane, the $\vec{e}_{L,y}$-axis in the $\vec{y}-\vec{z}$ plane, and the angle $\phi_L$ increasing counter-clockwise from the $\vec{e}_{L,x}$-axis. The angle $\xi$ is the angle between the vertical and the laser propagation axis as seen from the altitude $H$, i.e.:
\begin{equation}
\tan\xi = \frac{R_L-k}{H}
\end{equation}

The altitude $h_L$ of the sky element located at ($z_L$;$\rho_L$;$\phi_L$) can be written as:
 \begin{eqnarray}
 h_L &=& H - z_L \cos\xi  - y_B \sin\xi \nonumber\\
        &=& H - z_L \cos\xi  -\rho_L \sin\xi \sin\phi_L
 \end{eqnarray}
 so that the opening angle $\eta$ between the center of the MUSE field-of-view and the location \textit{on-the-sky} of the said element is, as seen from the location ($x_m$;$y_m$) on the primary mirror:
 \begin{equation}
 \eta = \frac{\pi}{2} - \arctan\left(\frac{h_L}{\sqrt{dx^2+dy^2}}\right)
 \end{equation}
 with:
 \begin{eqnarray}
 dx &=& x_m - x_L\\
 dy &=& y_m - y_L
 \end{eqnarray}

\begin{figure*}[htb!]
\centerline{\includegraphics[width=\textwidth]{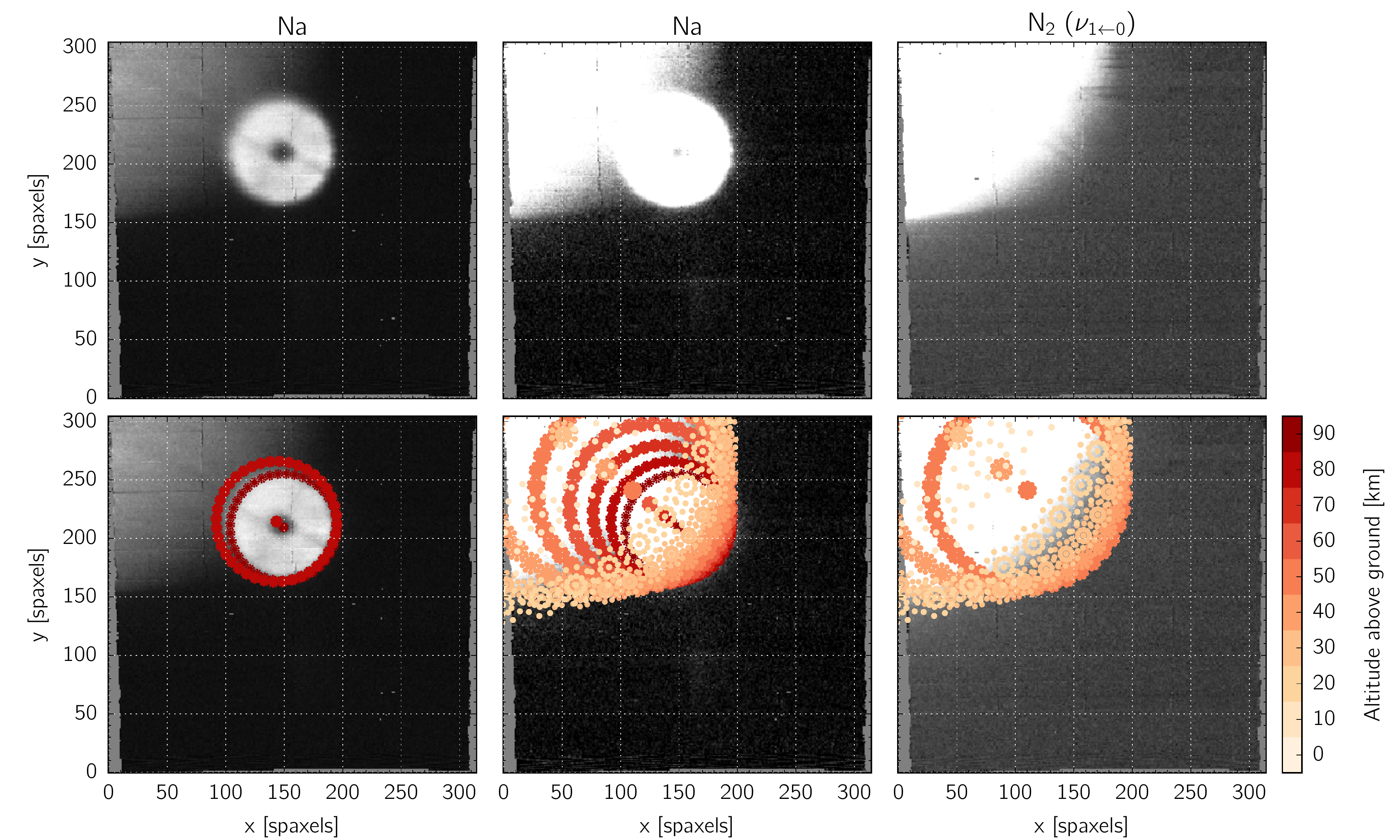}}
\caption{Top: summed MUSE slices within 4\,\AA\ from $\lambda_\mathrm{4LGSF}$ (left; middle: contrast enhanced) and 6826.73\,\AA\ ($\equiv$N$_2$ ($\nu_{1\leftarrow0}$); right) for an observation acquired with the LGS purposely located within the MUSE field-of-view. Bottom: idem, but with our modelled footprint of the laser beam as a function of the altitude overlaid (left: 80\,km \& 90\,km -- middle: 0\,km$\rightarrow$90\,km -- right: 10\,km$\rightarrow$50\,km). The out-of-focus LGS donut size is well reproduced by the model, with a diffuse extension to the top-left matching the $\sim$85\,km layer; and consistent with the typical thickness of a few km of the sodium layer. The Rayleigh and Raman beams are best matched by the model out to altitudes of $30\pm5$\,km above ground. Each star-like pattern in the modelled footprint of the 4LGSF beam (at a given altitude) is associated with one of our chosen sampling locations on the primary mirror -- which taken altogether trace the spatial extent of the un-resolved beam. A movie of the modeled beam footprint sampling the altitude in steps of 1\,km is available as supplementary material to the published article (in open access at \url{http://link.aps.org/supplemental/10.1103/PhysRevX.7.021044}).}\label{fig:model}
\end{figure*}

\clearpage

 Finally, this can be translated to coordinates in the MUSE frame (in radians):
 \begin{eqnarray}
 x_{MUSE} &=& \eta \cos\delta\\
 y_{MUSE} &=& \eta \sin\delta
 \end{eqnarray}
 with $\delta$ the azimuth (measured in the mirror plane) between the location ($x_m$;$y_m$) and the projected chunk of atmosphere at ($x_L$;$y_L$), which is:
 \begin{equation}
 \tan\delta = \frac{dy}{dx}
 \end{equation}
 
\begin{figure}[htb!]
\centerline{\includegraphics[scale=0.5]{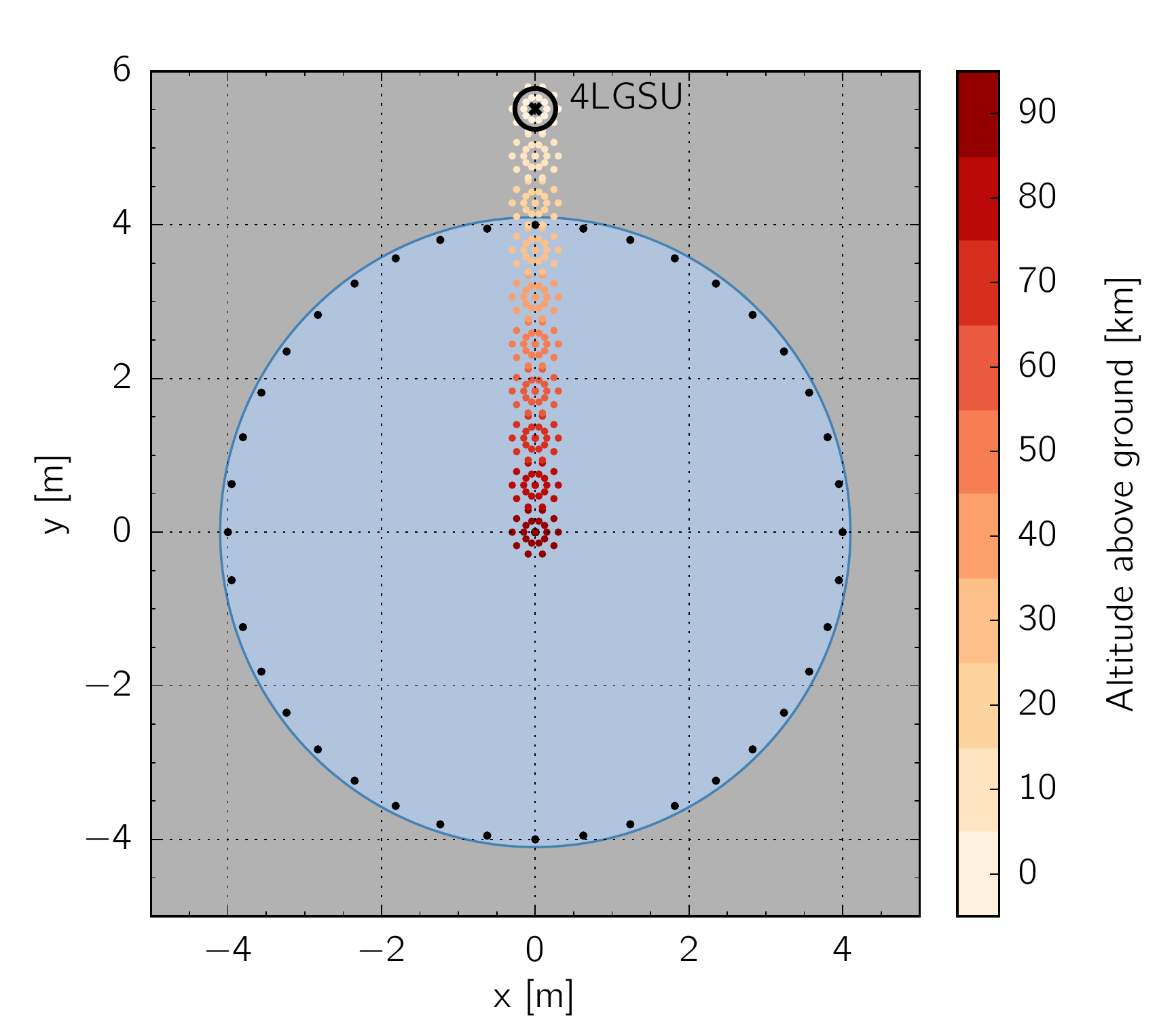}}
\caption{Sampling of the primary mirror plane in our 4LGSF+UT4 model, resulting in the sampling of the reconstructed laser beam images shown in Fig.~\ref{fig:model}. For clarity, the mirror is only sampled using its center and 60 locations on its outer-rim.}\label{fig:mirror}
\end{figure}

The prediction of this model -- sampling the beam in steps of 10\,km -- is presented in Fig.~\ref{fig:model} (bottom). The associated MUSE observation was performed at an altitude of 82$^{\circ}$, so that the beam altitudes $h_L$ computed via our model are off by only $\cos(90^{\circ}-82^{\circ})\equiv1\%$.  The size of the out-of-focus LGS \textit{donut} of 18 arcseconds in diameter is accurately reproduced. For clarity, we sampled the primary mirror with only its center point and 60 locations on the outer edge of the mirror (see Fig.~\ref{fig:mirror}), giving rise to the specific ring-like patterns visible in Fig.~\ref{fig:model}. We note that to match the orientation of the beam in the observation, we rotated our final model accounting for the value of the parallactic angle $\eta$ at the time of the observation. The Rayleigh (and Raman) beam is best matched by the model out to $30\pm5$\,km above ground (bottom right panel). For higher altitudes, we do not detect any evident sign of emission. For completeness, a movie illustrating the prediction of our model -- sampling the full laser beam at 1\,km intervals -- is made available online as supplementary material to the published article; in open access at \url{http://link.aps.org/supplemental/10.1103/PhysRevX.7.021044}.

\newpage
$ $
\newpage


%

\end{document}